\documentclass[11pt]{article}
\usepackage[T1]{fontenc}
\usepackage[utf8]{inputenc}
\usepackage{lmodern,microtype}
\usepackage[margin=1in]{geometry}
\usepackage{amsmath,amssymb,amsthm,mathtools}
\usepackage[colorlinks=true,linkcolor=blue!55!black,citecolor=blue!55!black,urlcolor=blue!55!black]{hyperref}
\usepackage{xcolor}
\hypersetup{pdftitle={Optimal mixing of random-to-random shuffling on linear extensions},pdfauthor={Haoran Wang},pdfsubject={Entropy contraction and mixing for the AST shuffle}}
\newtheorem{theorem}{Theorem}[section]
\newtheorem{lemma}[theorem]{Lemma}

\newtheorem{corollary}[theorem]{Corollary}
\theoremstyle{remark}
\DeclareMathOperator{\Ent}{Ent}
\DeclareMathOperator{\Var}{Var}
\DeclareMathOperator{\gap}{gap}
\DeclareMathOperator{\vol}{vol}
\DeclareMathOperator{\tr}{tr}
\newcommand{\E}{\mathbb E}
\newcommand{\R}{\mathbb R}
\newcommand{\cL}{\mathcal L}
\newcommand{\cO}{\mathcal O}
\newcommand{\cE}{\mathcal E}
\newcommand{\dd}{\,d}
\newcommand{\TV}{\mathrm{TV}}
\numberwithin{equation}{section}
\title{Optimal mixing of random-to-random shuffling\protect\\on linear extensions}
\author{Haoran Wang\\\small\href{mailto:whr.hrwang@gmail.com}{whr.hrwang@gmail.com}}
\date{}

\begin{document}
\maketitle

\begin{abstract}
Uniform sampling of linear extensions is a basic problem in randomized algorithms and the study of finite posets. We prove that the random-to-random shuffle of Ayyer, Schilling, and Thi\'ery mixes in $O(n\log n+n\log(1/\varepsilon))$ steps on every $n$-element poset, establishing their conjectured mixing order. With constant-time order queries, this gives an $\varepsilon$-approximate uniform sampler using $O(n^2(\log n+\log(1/\varepsilon)))$ elementary operations after preprocessing. The proof establishes relative-entropy contraction by a factor $1-1/n$ for the promotion half-step. Consequently, the uniform extended-promotion chain has the same mixing order. We represent complete promotion scans as reflected coordinate paths in the order polytope and prove that a convex potential with gradients in the same polytope has diagonal Hessian mass at most one along each path. Applying this estimate to an optimal transport potential yields a sharp entropy factorization for every finite-entropy input. Linearization also gives the spectral-gap lower bound $(n+1)/n^2$.
\end{abstract}

\section{Introduction}

Sampling a linear extension is a basic problem for finite partially ordered sets; Chan and Pak survey its combinatorial and computational context \cite{CP25}. A local way to move between extensions is to exchange two adjacent incomparable elements. Bubley and Dyer proved an $O(n^3\log n)$ mixing bound for a weighted adjacent-swap chain by path coupling \cite{BD99}. Wilson obtained the same order for the original Karzanov--Khachiyan adjacent-swap chain \cite{Wil04}. Ayyer, Schilling, and Thi\'ery introduced a random-to-random shuffle that groups adjacent swaps into insertion scans \cite{AST17}. They conjectured that this shuffle mixes in $O(n\log n)$ steps for every $n$-element poset. We prove this bound.

Let $P$ be a poset with $n\ge2$ elements. Write $\cL(P)$ for its linear extensions, viewed as words $x=x_1\cdots x_n$, and let $\pi$ be the uniform probability measure on $\cL(P)$. The map $\tau_i$ exchanges $x_i$ and $x_{i+1}$ when they are incomparable in $P$, and otherwise leaves the word fixed. One step of the random-to-random chain chooses $i,j\in\{1,\ldots,n\}$ independently and uniformly. If $i<j$, it applies $\tau_i,\tau_{i+1},\ldots,\tau_{j-1}$ in this chronological order. If $i>j$, it applies $\tau_{i-1},\tau_{i-2},\ldots,\tau_j$. If $i=j$, it does nothing. Denote its transition matrix by $M$. Time throughout the paper counts these complete updates.

The half-step is also useful. Let $U_j$ apply $\tau_{j-1},\ldots,\tau_1$ in chronological order, with $U_1$ the identity, and set
\begin{equation}\label{eq:K}
 (Kg)(x)=\frac1n\sum_{j=1}^n g(U_jx),
 \qquad
 (K^*g)(y)=\frac1n\sum_{j=1}^n g(U_j^{-1}y).
\end{equation}
Each $U_j$ is a bijection of $\cL(P)$, so the second operator is the adjoint of the first in $L^2(\pi)$. Both preserve $\pi$. The distinction between entropy contraction for a kernel and for its product with its adjoint is developed systematically by Caputo, Chen, Gu, and Polyanskiy \cite{CCGP25}. Cancellation of the overlapping scans gives
\begin{equation}\label{eq:factorM}
 M=KK^*.
\end{equation}
The corresponding state update is $U_i$ followed by $U_j^{-1}$; the overlapping adjacent involutions cancel in pairs.

For probability measures $\mu$ and $\pi$ on a finite set, write $D(\mu\Vert\pi)=\sum_x\mu(x)\log(\mu(x)/\pi(x))$, using natural logarithms and the convention $0\log0=0$.

\begin{theorem}\label{thm:main}
For every finite $n$-element poset $P$ and every probability measure $\mu$ on $\cL(P)$,
\begin{equation}\label{eq:half-contraction}
 D(\mu K\Vert\pi)\le\left(1-\frac1n\right)D(\mu\Vert\pi).
\end{equation}
Consequently, for every integer $t\ge0$,
\begin{equation}\label{eq:full-contraction}
 D(\mu M^t\Vert\pi)\le\left(1-\frac1n\right)^tD(\mu\Vert\pi).
\end{equation}
If $N=|\cL(P)|\ge2$ and $0<\varepsilon\le1/4$, then
\begin{equation}\label{eq:mix-explicit}
 t_{\mathrm{mix}}(\varepsilon)
 \le\left\lceil
 \frac{\log\bigl(\log N/(2\varepsilon^2)\bigr)}{-\log(1-1/n)}
 \right\rceil.
\end{equation}
In particular, $t_{\mathrm{mix}}(\varepsilon)=O(n\log n+n\log(1/\varepsilon))$, uniformly over $P$.
\end{theorem}

Here $t_{\mathrm{mix}}(\varepsilon)$ is the least $t$ for which $\max_x\|M^t(x,\cdot)-\pi\|_{\TV}\le\varepsilon$. For another kernel $A$ preserving $\pi$, write $t_{\mathrm{mix}}^A(\varepsilon)$ for the same quantity with $A$ in place of $M$. When $N=1$, this mixing time is zero. For an antichain, $M$ is the usual random-to-random card shuffle. Its mixing order is $n\log n$; Bernstein and Nestoridi proved a sharper cutoff theorem \cite{BN19}. Thus the order in Theorem~\ref{thm:main} is optimal in the worst case over posets.

Let $\rho$ be uniform measure on the order polytope
\begin{equation}\label{eq:polytope}
 \cO(P)=\{t\in[0,1]^P:t_a\le t_b\text{ whenever }a<_P b\}.
\end{equation}
Its equal-volume chambers are indexed by the linear extensions \cite{Sta86}. Marking one coordinate and running its promotion scan produces an exact disintegration of $\rho$. The conditional paths move in coordinate directions. At an allowed swap they continue straight, while at a blocked swap they turn from $e_a$ to $e_b$ with $a<_P b$.

Suppose $\psi$ is convex and all its subgradients lie in $\cO(P)$. Along such a turn, the directional slope can only increase, because each subgradient $p$ satisfies $p_a\le p_b$. All slopes lie between zero and one. Therefore the mass of the diagonal Hessian along a complete path is at most one. This is the geometric estimate proved in Section~\ref{sec:transport}.

For a positive density $F$ on $\cO(P)$, choose $\psi$ to transport $F\rho$ to $\rho$. Brenier's theorem supplies the convex potential \cite{Bre91}, and McCann's nonsmooth change-of-variables theorem gives $\det D_A^2\psi=F$ almost everywhere \cite{McC97}; here $D_A^2$ denotes the Alexandrov Hessian. Hadamard's inequality bounds $\log F$ by the sum of the logarithms of the diagonal Hessian entries. The pathwise mass estimate then allows a one-dimensional relative-entropy inequality on every path. Averaging proves a sharp continuous entropy factorization. Finally, discarding the auxiliary spacings gives \eqref{eq:half-contraction}.

Transport proofs of geometric inequalities have a long history, including Barthe's proof of the reverse Brascamp--Lieb inequality \cite{Bar98}. Cordero-Erausquin and Ledoux develop related geometric methods for entropy inequalities \cite{CL10}. The new estimate here is the unit bound on diagonal Hessian mass along a complete reflected promotion path. This bound allows the transport Jacobian to be combined with a one-dimensional entropy inequality on each path.

Entropy factorization for dependent variables is a general route to mixing estimates; see, for example, Caputo, Menz, and Tetali \cite{CMT15}. For an antichain the order polytope is a cube and our paths are coordinate fibers, so the continuous inequality reduces to product entropy tensorization, or the relative-entropy form of Han's inequality \cite{Han78}. For a general poset, the reflected paths supply a factorization with the same coefficient $n$.

For comparison with the usual promotion convention, let $\partial_j$ apply $\tau_j,\tau_{j+1},\ldots,\tau_{n-1}$ in chronological order, and let $Q$ choose $j$ uniformly from $\{1,\ldots,n\}$. These are the extended-promotion updates of Ayyer, Klee, and Schilling \cite{AKS14a}. Reversing a word and passing to the dual poset identifies $Q$ with the half-step $K$ on that dual poset.

\begin{corollary}\label{cor:promotion}
For every finite $n$-element poset, every initial law $\mu$, and every integer $t\ge0$,
\begin{equation}\label{eq:promotion-contraction}
 D(\mu Q^t\Vert\pi)\le\left(1-\frac1n\right)^tD(\mu\Vert\pi).
\end{equation}
Both $Q$ and $K$ have mixing time $O(n\log n+n\log(1/\varepsilon))$.
\end{corollary}

The mixing bound has a direct sampling consequence. We count operations on a word RAM with $O(\log n)$-bit indices and unit-cost access to independent uniform indices in $\{1,\ldots,n\}$. An order query tests the transitive relation $a<_P b$.

\begin{corollary}\label{cor:sampling}
Given an $n$-element poset through a constant-time order-query oracle and an initial linear extension, one can generate an extension within total variation distance $\varepsilon$ of uniform using $O(n^2(\log n+\log(1/\varepsilon)))$ elementary operations. If the input is instead an acyclic directed graph with $n$ vertices and $m$ edges, a transitive-closure table and an initial extension can be constructed in $O(n(n+m))$ time and $O(n^2)$ space. The same sampling bound then holds after this preprocessing.
\end{corollary}

For fixed accuracy, Corollary~\ref{cor:sampling} gives an operation count of $O(n^2\log n)$ after preprocessing. The adjacent-swap samplers analyzed by Bubley and Dyer and by Wilson use $O(n^3\log n)$ operations \cite{BD99,Wil04}. Huber's exact uniform sampler has expected running time $O(n^3\log n)$ \cite{Hub06}; his later work reduces the random-bit cost while retaining this order of operation count \cite{Hub25}. Leake, Lindberg, and Oveis Gharan study weighted linear-extension sampling through local spectral expansion of path complexes \cite{LLO25}.

A standard counting-from-sampling reduction also applies \cite{JVV86}. Appendix~\ref{app:counting} records the resulting counting bound and its proof, including the error from approximate samples.

The proof also yields a finite-size spectral estimate. For a real function $g$ on $\cL(P)$, let
\begin{equation}\label{eq:energy}
 \cE_M(g,g)=\langle g,(I-M)g\rangle_\pi.
\end{equation}
Since $M=KK^*$, it is self-adjoint and positive semidefinite. Its spectral gap is the infimum of $\cE_M(g,g)/\Var_\pi(g)$ over nonconstant $g$.

\begin{corollary}\label{cor:gap}
If $|\cL(P)|\ge2$, then
\begin{equation}\label{eq:gap}
 \gap(M)\ge\frac{n+1}{n^2}.
\end{equation}
\end{corollary}

Ayyer, Schilling, and Thi\'ery conjectured the stronger bound $(n+2)/n^2$, with equality precisely for disconnected posets, and derived an $O(n^2\log n)$ mixing bound conditional on this conjecture \cite{AST17}. The spectral correction in Corollary~\ref{cor:gap} comes from retaining auxiliary spacings before taking a variance expansion.

For the promotion chain, Ayyer, Klee, and Schilling obtained a cubic mixing bound for rooted forests with uniform element weights \cite{AKS14b}. Poznanovi\'c and Stasikelis obtained a quadratic bound for posets formed as an ordinal sum of a rooted forest and a ladder \cite{PS18}. Corollary~\ref{cor:promotion} gives the $O(n\log n)$ bound for every finite poset when the choice is uniform. Rhodes and Schilling obtained an $O(n\log n)$ mixing bound for a semigroup-based variant of the promotion chain \cite{RS22}.

For the ordinary random-to-random shuffle, Dieker and Saliola computed the spectrum and eigenspaces \cite{DS18}. On a general poset, the allowed-swap involutions can fail the symmetric-group braid relations \cite{AST17}. The reflected-path argument uses the order constraints directly.

Sections~\ref{sec:fibers} and~\ref{sec:transport} develop the continuous representation and entropy inequality. Section~\ref{sec:discrete} proves the discrete contraction and sampling bound. Section~\ref{sec:consequences} gives the promotion and continuous-time consequences, and Section~\ref{sec:spectral} derives the spectral estimates. Appendices~\ref{app:deficit} and~\ref{app:counting} contain the entropy deficit and the counting application.

\section{Marked coordinates and complete insertion paths}\label{sec:fibers}

For a probability measure $\nu$ and a nonnegative integrable function $h$ with finite entropy, write
\begin{equation}\label{eq:entropy}
 \Ent_\nu(h)=\int h\log h\dd\nu
 -\left(\int h\dd\nu\right)\log\left(\int h\dd\nu\right).
\end{equation}
In particular, $\Ent_\pi(d\mu/d\pi)=D(\mu\Vert\pi)$. The homogeneous form accommodates the varying masses of conditional functions along the paths.

The polytope $\cO(P)$ has nonempty interior in $\R^P$. Apart from coordinate ties, a point $t$ has a unique representation
\begin{equation}\label{eq:chambers}
 t_{x_1}=s_1<\cdots<s_n=t_{x_n},
 \qquad x\in\cL(P),\quad 0<s_1<\cdots<s_n<1.
\end{equation}
Each such chamber has volume $1/n!$. Thus $\vol(\cO(P))=N/n!$. Under $\rho$, the word $X$ is uniform on $\cL(P)$ and independent of $S=(S_1,\ldots,S_n)$. The latter has density $n!$ on the ordered simplex, which is the law of the order statistics of $n$ independent uniform points.

Choose a label $Z\in P$ uniformly and independently of $t$. If $Z=x_i$, define
\begin{equation}\label{eq:markmap}
 y=U_ix,\qquad q=(s_1,\ldots,s_{i-1},s_{i+1},\ldots,s_n),\qquad u=s_i.
\end{equation}
Let $\sigma_{n-1}$ denote the ordered law of $n-1$ independent uniform points in $[0,1]$.

\begin{lemma}\label{lem:mark}
Outside null sets, \eqref{eq:markmap} is a bijection from marked points $(t,Z)$ to triples $(y,q,u)$. Its image law is
\begin{equation}\label{eq:productlaw}
 \pi(dy)\,\sigma_{n-1}(dq)\,du.
\end{equation}
\end{lemma}

\begin{proof}
Given $(y,q,u)$, let $j$ be the rank of $u$ among the $n$ distinct points in $q\cup\{u\}$. Form the word $x=U_j^{-1}y$, assign these $n$ points to its labels in increasing order, and mark the label $x_j$. The resulting marked point has image $(y,q,u)$, because applying $U_j$ to $x$ recovers $y$. Conversely, the rank and the word are determined by any preimage, so this construction is unique.

Let $\mathcal S_n=\{0<s_1<\cdots<s_n<1\}$. On the piece with fixed word $x$ and mark $x_i$, the change of variables is
\[
 s_k=\begin{cases}q_k,&k<i,\\u,&k=i,\\q_{k-1},&k>i,\end{cases}
 \qquad q_{i-1}<u<q_i,
\]
with absolute Jacobian one. Thus for any bounded measurable test function $h$ of the output triple,
\begin{align*}
 \E h(Y,Q,U)
 &=\frac{n!}{Nn}\sum_{x\in\cL(P)}\sum_{i=1}^n
   \int_{\mathcal S_n}h(U_ix,s_{-i},s_i)\,ds\\
 &=\frac{(n-1)!}{N}\sum_{y\in\cL(P)}
   \int_{\mathcal S_{n-1}}\sum_{i=1}^n\int_{q_{i-1}}^{q_i}h(y,q,u)\,du\,dq\\
 &=\int h(y,q,u)\,\pi(dy)\sigma_{n-1}(dq)\,du.
\end{align*}
In the second line we used that each $U_i$ is a bijection. This proves the stated product law.
\end{proof}

Write $\Gamma_{y,q}(u)$ for the unmarked point obtained by this inverse construction. Put $q_0=0$ and $q_n=1$. For $q_{j-1}<u<q_j$, set
\begin{equation}\label{eq:carrier}
 x^{(j)}=U_j^{-1}y,\qquad z_j=x^{(j)}_j.
\end{equation}
Then $\Gamma_{y,q}$ is affine on this interval, with derivative $e_{z_j}$. Indeed, only the inserted value $u$ varies; every retained value in $q$ stays fixed. The label $z_j$ is the moving coordinate on that segment.

The relation $U_{j+1}^{-1}=\tau_j\circ U_j^{-1}$ gives $x^{(j+1)}=\tau_jx^{(j)}$. Let $a,b$ be the entries in positions $j,j+1$ of $x^{(j)}$. If $a$ and $b$ are incomparable, $\tau_j$ exchanges them. Just before the collision, the value of coordinate $a$ is $u$ and that of coordinate $b$ is $q_j$. Just after it, their values are still $u$ and $q_j$, respectively. Thus the same label $a$ continues to move, and the geometric path is straight through the chamber wall.

If $a$ and $b$ are comparable, their order in the extension forces $a<_P b$. The word is unchanged. Before the collision coordinate $a$ moves towards the fixed coordinate $b$; afterwards coordinate $b$ moves away from the fixed coordinate $a$. At $u=q_j$ the two coordinates agree. Hence the path is continuous and its direction changes from $e_a$ to $e_b$.

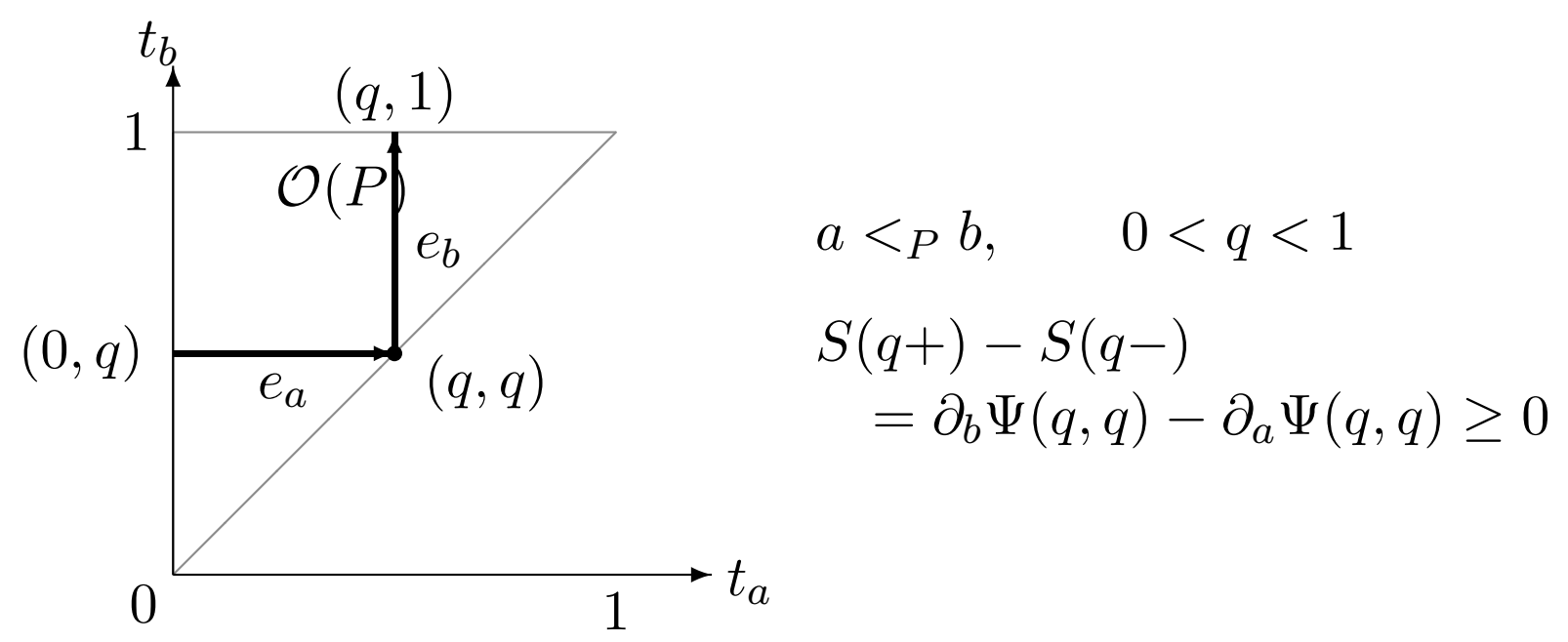
\begin{figure}[t]
\centering
\setlength{\unitlength}{1cm}
\begin{picture}(10,4.45)
\color{black!45}
\put(1,0.6){\line(0,1){3}}
\put(1,3.6){\line(1,0){3}}
\put(1,0.6){\line(1,1){3}}
\color{black}
\put(1,0.6){\vector(1,0){3.65}}
\put(1,0.6){\vector(0,1){3.45}}
\put(4.75,0.55){\makebox(0,0)[l]{$t_a$}}
\put(0.9,4.2){\makebox(0,0){$t_b$}}
\put(0.8,0.4){\makebox(0,0){$0$}}
\put(4,0.35){\makebox(0,0){$1$}}
\put(0.75,3.6){\makebox(0,0){$1$}}
\put(2.15,3.2){\makebox(0,0){$\mathcal O(P)$}}
\linethickness{1.2pt}
\put(1,2.1){\vector(1,0){1.5}}
\put(2.5,2.1){\vector(0,1){1.5}}
\put(2.5,2.1){\circle*{0.09}}
\put(0.8,2.1){\makebox(0,0)[r]{$(0,q)$}}
\put(2.5,3.85){\makebox(0,0){$(q,1)$}}
\put(2.7,1.9){\makebox(0,0)[l]{$(q,q)$}}
\put(1.75,1.85){\makebox(0,0){$e_a$}}
\put(2.8,2.8){\makebox(0,0){$e_b$}}
\put(5.35,2.9){\makebox(0,0)[l]{$a<_P b,\qquad 0<q<1$}}
\put(5.35,2.15){\makebox(0,0)[l]{$S(q+)-S(q-)$}}
\put(5.35,1.65){\makebox(0,0)[l]{$\quad=\partial_b\Psi(q,q)-\partial_a\Psi(q,q)\ge0$}}
\end{picture}
\caption{A complete reflected path for the two-element chain $a<_P b$. The path moves first in coordinate $a$ and then in coordinate $b$. If $\nabla\Psi$ lies in the same order polytope, the directional slope has a nonnegative jump at the turn.}
\label{fig:reflected-path}
\end{figure}

For example, on the chain $a<_P b$ with one retained value $q$, the path is
\[
 \Gamma(u)=
 \begin{cases}
 (u,q),&0\le u\le q,\\
 (q,u),&q\le u\le1.
 \end{cases}
\]
Figure~\ref{fig:reflected-path} illustrates this turn.

On the two-element antichain, the coordinate carrying $u$ continues straight instead. At every join with distinct retained values, the path either continues in the same coordinate direction or turns from $e_a$ to $e_b$ with $a<_P b$. Distinct retained values ensure that collisions occur one at a time, and this holds for $\sigma_{n-1}$-almost every $q$.

For nonnegative integrable $F$ on $\cO(P)$, define
\begin{equation}\label{eq:H}
 (HF)(y,q)=\int_0^1 F(\Gamma_{y,q}(u))\,du,
 \qquad
 \Delta(F)=\E_{y,q}\Ent_{du}(F\circ\Gamma_{y,q}).
\end{equation}
The expectation here and below uses $\pi\otimes\sigma_{n-1}$. Lemma~\ref{lem:mark} implies the averaging identity
\begin{equation}\label{eq:marked-average}
 \E_{y,q}\int_0^1 b(\Gamma_{y,q}(u),z(u))\,du
 =\frac1n\int\sum_{z\in P} b(t,z)\dd\rho(t),
\end{equation}
for every integrable function $b$ of a marked point. Here $z(u)=z_j$ on the $j$th segment. In particular, $\E HF=\rho(F)$.

The entropy chain rule takes the exact form
\begin{equation}\label{eq:chainrule}
 \Delta(F)=\Ent_\rho(F)-\Ent_{\pi\otimes\sigma_{n-1}}(HF).
\end{equation}
For a density $F$, conditional entropy on a path is
\[
 \Ent_{du}(F\circ\Gamma_{y,q})
 =\int_0^1 F(\Gamma_{y,q}(u))\log F(\Gamma_{y,q}(u))\,du
 -(HF)(y,q)\log(HF)(y,q).
\]
Averaging the first term by Lemma~\ref{lem:mark} gives $\rho(F\log F)$, while the second gives the entropy of the conditional expectation. More generally, if $m=\rho(F)$, then
\[
 \Delta(F)=\rho(F\log F)-\E[(HF)\log(HF)]
 =[\rho(F\log F)-m\log m]-[\E[(HF)\log(HF)]-m\log m],
\]
which proves \eqref{eq:chainrule}. These calculations first apply to bounded functions. For finite-entropy inputs, conditional Jensen bounds the positive part of $(HF)\log(HF)$ by the conditional mean of the positive part of $F\log F$, and the negative parts are bounded below by $-1/e$. Thus all the terms remain integrable.

Our main continuous inequality is
\begin{equation}\label{eq:continuous-target}
 \Ent_\rho(F)\le n\Delta(F).
\end{equation}
The retained values $q$ are essential in this factorization. Their spacing moments yield the finite-size correction in Corollary~\ref{cor:gap}.

\section{Convex transport and entropy factorization}\label{sec:transport}

The order constraints control how a convex potential changes along a reflected path. At a blocked swap the path turns in the direction of the larger element. If the gradient satisfies the same order constraints, its directional slope can only increase at that turn. Since every gradient coordinate lies in $[0,1]$, this bounds the total continuous increase of the slope by one.

\begin{lemma}\label{lem:smoothmass}
Let $\Psi$ be a smooth convex function on a neighborhood of $\cO(P)$, with $\nabla\Psi(t)\in\cO(P)$ for every $t\in\cO(P)$. For every path with distinct retained values,
\begin{equation}\label{eq:mass}
 \int_0^1\Psi_{z(u)z(u)}(\Gamma_{y,q}(u))\,du\le1.
\end{equation}
\end{lemma}

\begin{proof}
On each straight segment define the directional slope
\[
 S(u)=\partial_{z(u)}\Psi(\Gamma_{y,q}(u)).
\]
Its ordinary derivative on that segment is $\Psi_{z(u)z(u)}(\Gamma_{y,q}(u))\ge0$. At a straight join the carrier is unchanged and $S$ is continuous. At a turn from $e_a$ to $e_b$, the common point lies on $t_a=t_b$ and $a<_P b$. Since $\nabla\Psi$ belongs to the order polytope, the jump is
\begin{equation}\label{eq:jump}
 S(q_j+)-S(q_j-)=\partial_b\Psi(\Gamma(q_j))-\partial_a\Psi(\Gamma(q_j))\ge0.
\end{equation}
Since $\nabla\Psi(t)\in\cO(P)\subset[0,1]^P$, we have $0\le S(u)\le1$ everywhere on each segment. In particular, $S(0+)\ge0$ and $S(1-)\le1$. Adding the increments over all segments yields
\begin{align*}
 \int_0^1\Psi_{z(u)z(u)}(\Gamma(u))\,du
 &=S(1-)-S(0+)
   -\sum_{\text{turns }a\to b}(\partial_b\Psi-\partial_a\Psi)(\Gamma(q_j))\\
 &\le1.
\end{align*}
\end{proof}

Thus $\Psi\circ\Gamma$ is convex, with slopes in $[0,1]$. Positive corner jumps reduce the mass available to the ordinary second derivatives. The next lemma passes this bound to the almost-everywhere Hessian of a convex potential.

\begin{lemma}\label{lem:nonsmoothmass}
Let $\Psi$ be a finite convex function on $\R^P$ such that every subgradient of $\Psi$ lies in $\cO(P)$. Let $A$ denote its Alexandrov Hessian, defined almost everywhere. Then for almost every $(y,q)$,
\begin{equation}\label{eq:nonsmoothmass}
 \int_0^1 A_{z(u)z(u)}(\Gamma_{y,q}(u))\,du\le1.
\end{equation}
Consequently,
\begin{equation}\label{eq:tracemass}
 \int\tr A\dd\rho\le n.
\end{equation}
\end{lemma}

\begin{proof}
The distributional Hessian of a finite convex function is a locally finite positive semidefinite matrix of measures. Its absolutely continuous density agrees almost everywhere with the Alexandrov Hessian. Choose a nonnegative smooth mollifier $\eta_\delta$, supported in a ball of radius $\delta$, and put $\Psi_\delta=\Psi*\eta_\delta$. The subgradient assumption implies that $\Psi$ is globally Lipschitz. Its almost-everywhere gradient lies in the compact convex set $\cO(P)$, so averaging gives
\[
 \nabla\Psi_\delta(t)=\int\nabla\Psi(t-v)\eta_\delta(v)\,dv\in\cO(P)
\]
for every $t\in\R^P$. Therefore Lemma~\ref{lem:smoothmass} applies to $\Psi_\delta$.

Write the Lebesgue decomposition of the distributional Hessian as $D^2\Psi=A(t)\,dt+\mu$, where $\mu$ is a singular positive semidefinite matrix-valued measure. For every pair of coordinates,
\begin{equation}\label{eq:measure-domination}
 |\mu_{ij}|\le\tfrac12(\mu_{ii}+\mu_{jj})\le\tr\mu
\end{equation}
as measures, with $|\mu_{ij}|$ denoting total variation. Indeed, taking Radon--Nikodym derivatives with respect to $\lambda=\tr\mu$ gives a positive semidefinite matrix $Q$ almost everywhere, and $|Q_{ij}|\le\sqrt{Q_{ii}Q_{jj}}\le(Q_{ii}+Q_{jj})/2$. Integrating proves \eqref{eq:measure-domination}.

Choose the mollifiers in the scaled form $\eta_\delta(v)=\delta^{-n}\eta(v/\delta)$, with $\eta$ supported in the unit ball. The Lebesgue differentiation theorem for singular measures gives
\[
 \lim_{\delta\downarrow0}\frac{\lambda(B(t,\delta))}{\vol(B(t,\delta))}=0
 \quad\text{for Lebesgue-almost every }t.
\]
Consequently, \eqref{eq:measure-domination} yields
\[
 |(\mu_{ij}*\eta_\delta)(t)|
 \le\|\eta\|_\infty\delta^{-n}\lambda(B(t,\delta))\longrightarrow0
\]
at almost every point. Convolution of $A(t)\,dt$ converges to $A$ at its Lebesgue points, and therefore
\begin{equation}\label{eq:mollifyhess}
 D^2\Psi_\delta(t)\longrightarrow A(t)
 \quad\text{for Lebesgue-almost every }t.
\end{equation}

Fix a sequence $\delta\downarrow0$, and let $N_0$ be a Borel null set outside which \eqref{eq:mollifyhess} holds and $A$ is defined. By Lemma~\ref{lem:mark}, the image of $\pi(dy)\sigma_{n-1}(dq)\,du$ under $\Gamma_{y,q}(u)$ is $\rho$. Hence
\[
 \E_{y,q}\int_0^1\mathbf1_{N_0}(\Gamma_{y,q}(u))\,du=\rho(N_0)=0.
\]
Fubini's theorem now shows that, for almost every $(y,q)$, convergence holds for almost every path parameter $u$. All diagonal entries are nonnegative. Fatou's lemma and the smooth path bound give
\[
 \int_0^1 A_{z(u)z(u)}(\Gamma(u))\,du
 \le\liminf_{\delta\downarrow0}\int_0^1(\Psi_\delta)_{z(u)z(u)}(\Gamma(u))\,du\le1.
\]
Finally, average this inequality and use \eqref{eq:marked-average}. Its left side is $(1/n)\int\tr A\dd\rho$, proving \eqref{eq:tracemass}.
\end{proof}

\begin{theorem}\label{thm:continuous}
For every finite $n$-element poset and every nonnegative integrable function $F$ of finite entropy on its order polytope,
\begin{equation}\label{eq:continuous}
 \Ent_\rho(F)\le n\E_{y,q}\Ent_{du}(F\circ\Gamma_{y,q}).
\end{equation}
Equivalently,
\begin{equation}\label{eq:continuouschannel}
 \Ent_{\pi\otimes\sigma_{n-1}}(HF)
 \le\left(1-\frac1n\right)\Ent_\rho(F).
\end{equation}
For each poset, the coefficient $n$ in \eqref{eq:continuous} is optimal.
\end{theorem}

\begin{proof}

Let $r\ge0$ be bounded on $[0,1]$, and let $k>0$ almost everywhere satisfy $a=\int_0^1k\le1$. If $m=\int_0^1r>0$, the nonnegativity of relative entropy gives
\begin{equation}\label{eq:gibbs}
 \int_0^1r\log k
 =\Ent_{du}(r)-mD\left(\frac rm\middle\Vert\frac ka\right)+m\log a
 \le\Ent_{du}(r).
\end{equation}
Here the two normalized functions in $D$ are densities with respect to $du$. The formula holds whenever the left side is integrable, and the inequality remains valid if it equals $-\infty$. If $m=0$, both sides are zero. The condition $a\le1$ is precisely what gives the sign of the last term.

The zero function is immediate, so homogeneity allows us to assume $\rho(F)=1$. We first suppose that $0<c\le F\le C<\infty$. Apply quadratic optimal transport to the source $F\rho$ and target $\rho$. Brenier's theorem \cite{Bre91} provides a proper lower semicontinuous convex function $\psi:\R^P\to\R\cup\{+\infty\}$ whose gradient satisfies
\begin{equation}\label{eq:brenier}
 (\nabla\psi)_\#(F\rho)=\rho.
\end{equation}
The function is finite on a set of full source measure. Since $F\ge c>0$, this set has full Lebesgue measure in $\cO(P)$. Every point in the interior of $\cO(P)$ lies in the interior of a simplex with vertices in this full-measure set. Convexity then implies that $\psi$ is finite on the entire interior. In particular, it is continuous and locally Lipschitz there, and Alexandrov's theorem supplies an almost-everywhere Hessian $A=D_A^2\psi$.

Regard both transport densities as Lebesgue densities on $\R^P$. The target has density $\mathbf1_{\operatorname{int}\cO(P)}/\vol(\cO(P))$. McCann's monotone change-of-variables theorem \cite[Theorem~4.4 and Remark~4.5]{McC97}, also discussed in \cite{DPF14}, gives
\[
 \frac{F(t)}{\vol(\cO(P))}
 =\frac{\mathbf1_{\operatorname{int}\cO(P)}(\nabla\psi(t))}{\vol(\cO(P))}\det A(t)
 \quad\text{for }F\rho\text{-almost every }t.
\]
The boundary of the polytope is Lebesgue-null, so it has zero target measure. Equation~\eqref{eq:brenier} implies that $\nabla\psi(t)$ belongs to the interior for source-almost every $t$. The target-density factor in this identity is therefore $1/\vol(\cO(P))$. Cancelling the volume factors and using $F\ge c$ yields
\begin{equation}\label{eq:jacobian}
 \det A(t)=F(t)\qquad\rho\text{-almost everywhere}.
\end{equation}
Since $A$ is positive semidefinite and its determinant is positive, it is positive definite almost everywhere.

The map $\nabla\psi$ lies in $\cO(P)$ almost everywhere in the interior. We need a convex extension for which this slope constraint holds globally. Let $E$ be a full-measure set of differentiability points in the interior where the gradient belongs to $\cO(P)$, and define
\begin{equation}\label{eq:extension}
 \bar\psi(t)=\sup_{x\in E}
 \{\psi(x)+\nabla\psi(x)\cdot(t-x)\},\qquad t\in\R^P.
\end{equation}

All gradients on $E$ have norm at most $\sqrt n$. A convex function with this almost-everywhere gradient bound is $\sqrt n$-Lipschitz on its open convex domain. One can obtain the bound by integrating along almost every segment and then using continuity to pass to the remaining segments. Therefore $\psi$ extends continuously to the closed polytope and is bounded there. The affine functions in \eqref{eq:extension} have uniformly bounded slopes and intercepts, so their supremum is finite everywhere and globally Lipschitz.

Every plane in \eqref{eq:extension} supports $\psi$ on the interior. Thus $\bar\psi\le\psi$ there, with equality at every point of $E$. Density of $E$ and continuity imply equality throughout the interior. In particular, the almost-everywhere Hessian of $\bar\psi$ on the polytope is still $A$.

For the slope constraint, let $v\in\R^P$ and write $h(v)=\sup_{p\in\cO(P)}p\cdot v$. The supremum representation implies
\[
 \bar\psi(t+v)-\bar\psi(t)\le h(v).
\]
For any $p\in\partial\bar\psi(t)$, the subgradient inequality gives $p\cdot v\le h(v)$ for every $v$. The separating-hyperplane characterization of a closed convex set then yields $p\in\cO(P)$. Lemma~\ref{lem:nonsmoothmass} therefore applies to $\bar\psi$.

By \eqref{eq:tracemass}, $\int\tr A\dd\rho\le n$. Since $\log^+s\le s$ for $s>0$, each $\log^+A_{zz}$ is integrable. Hadamard's inequality and \eqref{eq:jacobian} give
\begin{equation}\label{eq:hadamard}
 \sum_{z\in P}\log A_{zz}\ge\log\det A=\log F.
\end{equation}
If $A_{zz}<1$, this implies
\[
 -\log A_{zz}\le\sum_{w\ne z}\log^+A_{ww}+|\log c|.
\]
Consequently, $F\log A_{zz}$ is integrable for every $z$, so the pathwise logarithmic inequalities can be averaged.

For almost every path put
\[
 r(u)=F(\Gamma_{y,q}(u)),\qquad
 k(u)=A_{z(u)z(u)}(\Gamma_{y,q}(u)).
\]
By marked disintegration, $k>0$ almost everywhere on almost every path. Lemma~\ref{lem:nonsmoothmass} gives $\int k\le1$, and averaging \eqref{eq:gibbs} yields
\begin{equation}\label{eq:logaverage}
 \frac1n\int F(t)\sum_{z\in P}\log A_{zz}(t)\dd\rho(t)
 \le\E_{y,q}\Ent_{du}(F\circ\Gamma_{y,q})=\Delta(F).
\end{equation}
On the other hand, \eqref{eq:jacobian} and \eqref{eq:hadamard} imply
\[
 \Ent_\rho(F)=\int F\log F\dd\rho
 =\int F\log\det A\dd\rho
 \le\int F\sum_{z\in P}\log A_{zz}\dd\rho.
\]
Together with \eqref{eq:logaverage}, this proves \eqref{eq:continuous} for densities bounded above and below by positive constants.

For a bounded nonnegative density $F$, use $F_\epsilon=(1-\epsilon)F+\epsilon$. The preceding result applies to $F_\epsilon$. Both $F_\epsilon$ and $HF_\epsilon$ are uniformly bounded, and $s\log s$ is continuous on bounded nonnegative intervals. Bounded convergence therefore gives convergence of their entropies. Passing to the limit in \eqref{eq:continuouschannel} proves the inequality for bounded densities. For an arbitrary density $F$ of finite entropy, let $c_L=\rho(F\wedge L)$ and $F_L=(F\wedge L)/c_L$, with $L\ge1$. Then $F_L\to F$ in $L^1(\rho)$ and $\Ent_\rho(F_L)\to\Ent_\rho(F)$, by truncation of the positive entropy part and $c_L\to1$. Conditional expectation is an $L^1$ contraction, so $HF_L\to HF$ in $L^1$. Lower semicontinuity of entropy gives
\[
 \Ent_{\pi\otimes\sigma_{n-1}}(HF)\le\liminf_{L\to\infty}\Ent_{\pi\otimes\sigma_{n-1}}(HF_L)\le\left(1-\frac1n\right)\Ent_\rho(F).
\]
The chain rule \eqref{eq:chainrule} extends to these inputs by conditional Jensen and integrability of the entropy terms. Homogeneity completes the assertion for arbitrary finite mass.

Take any bounded positive density $h$ on $[0,1]$ with $\Ent_{du}(h)>0$, and set
\begin{equation}\label{eq:productequality}
 F(t)=\prod_{z\in P}h(t_z).
\end{equation}
Since the expression is symmetric in the coordinate values, chamber decomposition gives
\[
 \int F\dd\rho
 =n!\int_{0<s_1<\cdots<s_n<1}\prod_{i=1}^nh(s_i)\,ds=1.
\]
The same symmetry, applied to $F\log F$, gives $\Ent_\rho(F)=n\Ent_{du}(h)$. Along a path the multiset of coordinate values is $\{q_1,\ldots,q_{n-1},u\}$. Consequently,
\[
 F(\Gamma_{y,q}(u))=h(u)\prod_{i=1}^{n-1}h(q_i).
\]
Homogeneity of entropy and the ordered-sample law of $q$ now give $\Delta(F)=\Ent_{du}(h)$. Thus equality holds in \eqref{eq:continuous} for this nonconstant $F$. This proves optimality of $n$ for every poset.
\end{proof}

\section{Compression to the discrete shuffle}\label{sec:discrete}

Let $f=d\mu/d\pi$. We apply the continuous theorem to the chamber function $F(t)=f(X(t))$, then average over the retained spacings.

\begin{proof}[Proof of Theorem~\ref{thm:main}]

Independence of $X$ and $S$ gives $\Ent_\rho(F)=\Ent_\pi(f)$. For fixed $(y,q)$, the word on the $j$th segment is $U_j^{-1}y$. Therefore
\begin{equation}\label{eq:spacings}
 (HF)(y,q)=\sum_{j=1}^n G_j f(U_j^{-1}y),\qquad G_j=q_j-q_{j-1}.
\end{equation}
The spacing vector $G$ has the uniform Dirichlet law $\operatorname{Dir}(1,\ldots,1)$, with $n$ components. It is independent of $y$ and satisfies $\E G_j=1/n$.

For each fixed $q$, the function $HF(\cdot,q)$ has mean one under $\pi$, since each $U_j^{-1}$ is a bijection. Hence the joint entropy in Theorem~\ref{thm:continuous} equals the mean over $G$ of the entropy in $y$. We obtain
\begin{equation}\label{eq:spacingentropy}
 \E_G\Ent_\pi\left(\sum_jG_jf\circ U_j^{-1}\right)
 \le\left(1-\frac1n\right)\Ent_\pi(f).
\end{equation}
Convexity of $s\log s$, followed by $\E G_j=1/n$, yields
\[
 \Ent_\pi(K^*f)
 =\Ent_\pi\left(\E_G\sum_jG_jf\circ U_j^{-1}\right)
 \le\E_G\Ent_\pi\left(\sum_jG_jf\circ U_j^{-1}\right).
\]
The density of the row measure $\mu K$ relative to $\pi$ is $K^*f$. Thus \eqref{eq:spacingentropy} proves \eqref{eq:half-contraction}.

To identify $KK^*$, start at a word $x$, choose $i,j$ independently and uniformly, and apply $U_i$ followed by $U_j^{-1}$. The chronological list of involutions is
\[
 \tau_{i-1},\ldots,\tau_1,\tau_1,\ldots,\tau_{j-1}.
\]
Consecutive equal involutions cancel. If $i<j$, the remaining list is $\tau_i,\ldots,\tau_{j-1}$. If $i>j$, it is $\tau_{i-1},\ldots,\tau_j$. This is the AST update, proving \eqref{eq:factorM}.

Both $K$ and $K^*$ preserve $\pi$, so relative entropy cannot increase under either kernel. Apply \eqref{eq:half-contraction} to the first update and entropy monotonicity to the second. Iteration gives \eqref{eq:full-contraction}.

The chain is irreducible: Given a target extension, fix its positions from left to right. If the next required element lies at a later position, all elements that it must pass are incomparable with it: they precede it in the current extension but follow it in the target extension. Allowed adjacent swaps therefore move it into place. Each such adjacent swap has positive probability under $M$. The choices $i=j$ give a positive holding probability, so the chain is also aperiodic.

Starting at $x$, the initial relative entropy is $D(\delta_x\Vert\pi)=\log N$. Pinsker's inequality and \eqref{eq:full-contraction} imply
\begin{equation}\label{eq:pinsker}
 \|M^t(x,\cdot)-\pi\|_{\TV}^2
 \le\frac12\left(1-\frac1n\right)^t\log N.
\end{equation}
Solving for $t$ proves \eqref{eq:mix-explicit}. Since $-\log(1-1/n)\ge1/n$ and $N\le n!$, it also gives the stated uniform asymptotic estimate.
\end{proof}

\begin{proof}[Proof of Corollary~\ref{cor:sampling}]
Represent a linear extension by an array of labels. Each allowed adjacent involution uses one order query and at most one array swap, both taking constant time in the stated model. A complete AST update performs at most $n-1$ such involutions, together with two uniform index choices. Theorem~\ref{thm:main} therefore gives the claimed number of elementary operations.

For a directed acyclic graph presentation, a graph search from each vertex computes its reachable set in $O(n(n+m))$ total time. Storing these sets as an $n\times n$ Boolean table gives constant-time order queries and uses $O(n^2)$ space. A topological ordering supplies the initial linear extension in $O(n+m)$ time. Thus the total running time, including this preprocessing, is
\[
 O\bigl(n(n+m)+n^2(\log n+\log(1/\varepsilon))\bigr).
\]
Boolean transitive-closure algorithms provide an alternative to the graph-search preprocessing.
\end{proof}

The coefficient $1/n$ in the entropy loss is also asymptotically optimal among bounds that hold for all posets and all inputs. On the $n$-element antichain, fix a label $a$ and let $\mu$ be uniform conditional on $a$ occupying the last position. Then $D(\mu\Vert\pi)=\log n$.

The random-to-top half-step chooses $a$ with probability $1/n$, in which case $a$ becomes first. Otherwise $a$ stays last. Conditional on either event, the other labels are uniformly ordered. Hence the output density relative to $\pi$ equals $1$ when $a$ is first, $n-1$ when it is last, and zero otherwise. Therefore
\[
 D(\mu K\Vert\pi)=\frac{n-1}{n}\log(n-1),
 \qquad
 D(\mu\Vert\pi)-D(\mu K\Vert\pi)=h_2(1/n),
\]
where $h_2(p)=-p\log p-(1-p)\log(1-p)$. It follows that
\[
 \frac{n\,[D(\mu\Vert\pi)-D(\mu K\Vert\pi)]}{D(\mu\Vert\pi)}
 =\frac{n h_2(1/n)}{\log n}\longrightarrow1.
\]
Thus no fixed constant larger than one can replace one in a uniform entropy-loss bound $D(\mu\Vert\pi)-D(\mu K\Vert\pi)\ge(c/n)D(\mu\Vert\pi)$ for all $n$.

\section{Promotion, position weights, and continuous time}\label{sec:consequences}

\begin{proof}[Proof of Corollary~\ref{cor:promotion}]
Let $P^{\mathrm{op}}$ be the dual poset, and let $R(x_1\cdots x_n)=x_n\cdots x_1$. This is a bijection from $\cL(P)$ to $\cL(P^{\mathrm{op}})$ preserving uniform measure. An adjacent involution at position $i$ becomes the involution at position $n-i$ after reversal. Thus the chronological scan $\tau_j,\ldots,\tau_{n-1}$ becomes $\tau_{n-j},\ldots,\tau_1$, which is $U_{n+1-j}$ on the dual poset. Uniform choice of $j$ is preserved, so $Q_P$ is conjugate to $K_{P^{\mathrm{op}}}$.

Apply Theorem~\ref{thm:main} to the dual poset and iterate its half-step contraction. This gives \eqref{eq:promotion-contraction}. Direct iteration on $P$ gives the same estimate for $K^t$. Starting from a state, the initial entropy is $\log|\cL(P)|\le\log(n!)$, so Pinsker's inequality gives both mixing bounds as in \eqref{eq:pinsker}.
\end{proof}

The half-step theorem also allows arbitrary probabilities on the scan positions, with the uniform measure remaining stationary.

\begin{corollary}\label{cor:position-weights}
Let $p_1,\ldots,p_n$ be probabilities, set $p_* =\min_j p_j$, and let $K_p$ choose $U_j$ with probability $p_j$. Then
\begin{equation}\label{eq:position-contraction}
 D(\mu K_p\Vert\pi)\le(1-p_*)D(\mu\Vert\pi).
\end{equation}
The same conclusion holds for $Q_p$, which chooses $\partial_j$ with probability $p_j$. If $p_*>0$ and $N=|\cL(P)|\ge2$, their mixing times satisfy
\[
 \max\{t_{\mathrm{mix}}^{K_p}(\varepsilon),t_{\mathrm{mix}}^{Q_p}(\varepsilon)\}
 \le\left\lceil\frac{\log(\log N/(2\varepsilon^2))}{-\log(1-p_*)}\right\rceil,
 \qquad 0<\varepsilon\le1/4.
\]
\end{corollary}

\begin{proof}
Each $U_j$ preserves $\pi$. Decompose the update as a mixture with weight $np_*$ on $K$ and weights $p_j-p_*$ on the individual bijections $U_j$. Convexity of relative entropy and invariance under each bijection yield
\begin{align*}
 D(\mu K_p\Vert\pi)
 &\le np_*D(\mu K\Vert\pi)+\sum_j(p_j-p_*)D(\mu U_j\Vert\pi)\\
 &\le\left[np_*\left(1-\frac1n\right)+1-np_*\right]D(\mu\Vert\pi)
 =(1-p_*)D(\mu\Vert\pi).
\end{align*}
For $Q_p$, reversal gives $K_{p'}$ on the dual poset, where $p'_j=p_{n+1-j}$ and $\min_jp'_j=p_*$. Iteration and Pinsker's inequality give the stated mixing bounds.
\end{proof}

Discrete entropy contraction also gives continuous-time decay and a modified logarithmic Sobolev inequality; see Caputo, Chen, Gu, and Polyanskiy for general comparisons \cite{CCGP25}.

\begin{corollary}\label{cor:continuous-time}
For the unit-rate generator $L=M-I$,
\begin{equation}\label{eq:continuous-time}
 D(\mu e^{tL}\Vert\pi)\le e^{-t/n}D(\mu\Vert\pi),\qquad t\ge0.
\end{equation}
Equivalently, its modified logarithmic Sobolev constant is at least $1/n$ under the convention
\begin{equation}\label{eq:mlsi}
 \langle f,(I-M)\log f\rangle_\pi\ge\frac1n\Ent_\pi(f),\qquad f>0.
\end{equation}
\end{corollary}

\begin{proof}
The semigroup is $e^{t(M-I)}=e^{-t}\sum_{k\ge0}(t^k/k!)M^k$. By convexity of relative entropy and \eqref{eq:full-contraction},
\[
 D(\mu e^{tL}\Vert\pi)
 \le e^{-t}\sum_{k\ge0}\frac{t^k}{k!}\left(1-\frac1n\right)^kD(\mu\Vert\pi)
 =e^{-t/n}D(\mu\Vert\pi).
\]
For a positive density $f$, differentiate at $t=0$. Self-adjointness of $M$ identifies the entropy derivative as $-\langle f,(I-M)\log f\rangle_\pi$, proving \eqref{eq:mlsi}. Homogeneity extends the inequality to every positive $f$. Conversely, applying \eqref{eq:mlsi} to the evolving density and integrating the resulting differential inequality gives \eqref{eq:continuous-time}.
\end{proof}

\section{Spectral consequences}\label{sec:spectral}

\subsection{The spectral gap}

For bounded real $B$ on the order polytope define
\[
 V(B)=\E_{y,q}\Var_{du}(B\circ\Gamma_{y,q}).
\]
The second-order expansion of homogeneous entropy is
\begin{equation}\label{eq:entropy-expansion}
 \Ent_\nu(1+\epsilon b)=\frac{\epsilon^2}{2}\Var_\nu(b)+O_b(\epsilon^3)
\end{equation}
for bounded $b$. Indeed, expand $(1+s)\log(1+s)=s+s^2/2+O(s^3)$ and subtract the same expansion at $s=\epsilon\nu(b)$. The remainder is uniform when $b$ is uniformly bounded. Applying \eqref{eq:entropy-expansion} to Theorem~\ref{thm:continuous} yields
\begin{equation}\label{eq:continuousvariance}
 \Var_\rho(B)\le n V(B).
\end{equation}

Take $B(t)=g(X(t))$, where $\pi(g)=0$. For fixed $y$, write $a_j(y)=g(U_j^{-1}y)$. The Dirichlet spacings satisfy
\begin{equation}\label{eq:dirichletmoments}
 \E G_j^2=\frac{2}{n(n+1)},\qquad
 \E G_jG_k=\frac{1}{n(n+1)}\quad(j\ne k).
\end{equation}
Consequently,
\begin{equation}\label{eq:squaremean}
 \E_G\left(\sum_jG_ja_j\right)^2
 =\frac{\sum_ja_j^2+(\sum_ja_j)^2}{n(n+1)}.
\end{equation}
The identities in \eqref{eq:dirichletmoments} follow by integrating monomials on the uniform simplex; they also follow from the beta marginals of a uniform Dirichlet vector.

Because the $U_j$ preserve $\pi$, averaging $\sum_j a_j^2$ gives $n\|g\|_2^2$. Also $\sum_j a_j=nK^*g$. The total variance identity and \eqref{eq:squaremean} now give
\begin{align}\label{eq:varianceenergy}
 V(B)
 &=\|g\|_2^2-\frac{1}{n+1}\|g\|_2^2
   -\frac{n}{n+1}\|K^*g\|_2^2\notag\\
 &=\frac{n}{n+1}\bigl(\|g\|_2^2-\|K^*g\|_2^2\bigr)
 =\frac{n}{n+1}\cE_M(g,g).
\end{align}
Combining \eqref{eq:continuousvariance} and \eqref{eq:varianceenergy} proves
\[
 \Var_\pi(g)\le\frac{n^2}{n+1}\cE_M(g,g),
\]
which is Corollary~\ref{cor:gap}.

\subsection{Ordinal sums}

An ordinal sum $P=P_1\oplus\cdots\oplus P_r$ places every element of $P_i$ below every element of $P_j$ for $i<j$. Its extensions are concatenations of extensions of the blocks. Thus $\pi=\pi_1\otimes\cdots\otimes\pi_r$. Put $m_i=|P_i|$, and let $M_i$ denote the AST chain within $P_i$.

\begin{corollary}\label{cor:ordinal}
For a nontrivial ordinal sum,
\begin{equation}\label{eq:ordinalcomparison}
 \gap(M_P)\ge\min_{i:\,|\cL(P_i)|>1}\frac{m_i}{n}\gap(M_i).
\end{equation}
In particular, if $m_* =\max\{m_i:|\cL(P_i)|>1\}$, then
\begin{equation}\label{eq:ordinalbound}
 \gap(M_P)\ge\frac1n\left(1+\frac1{m_*}\right).
\end{equation}
When $m_*\le n/2$, this is at least $(n+2)/n^2$.
\end{corollary}

\begin{proof}
For $v_j=g\circ U_j^{-1}$, the energy is the variance of $n$ vectors in $L^2(\pi)$:
\[
 \cE_{M_P}(g,g)=\frac1n\sum_j\|v_j\|_2^2
 -\left\|\frac1n\sum_jv_j\right\|_2^2.
\]
Group the indices $j$ by their ordinal block. The variance decomposition in this Hilbert space implies that the total variance is at least the weighted sum of the within-group variances, with weights $m_i/n$.

Within block $i$, write $j=m_1+\cdots+m_{i-1}+\ell$. The inverse scan $U_j^{-1}$ performs the same complete inverse scan on every earlier block, followed by the local inverse scan $U_{i,\ell}^{-1}$ on block $i$. Swaps at block boundaries are blocked. The common map on the earlier blocks is a measure-preserving bijection independent of $\ell$ and acts on different factors from $U_{i,\ell}^{-1}$. It can therefore be removed from the $L^2(\pi)$ norms and averages in the within-group variance. That variance is exactly the conditional energy for $M_i$. We obtain
\begin{equation}\label{eq:ordinalenergy}
 \cE_{M_P}(g,g)\ge
 \sum_i\frac{m_i}{n}\E\,\cE_{M_i}(g(\cdot,X_{-i}),g(\cdot,X_{-i})).
\end{equation}
Apply the spectral inequality in each nontrivial factor and then the product variance inequality
\[
 \Var_\pi(g)\le\sum_i\E\Var_{\pi_i}(g(\cdot,X_{-i})).
\]
This proves \eqref{eq:ordinalcomparison}. A factor with one extension has zero conditional variance and can be omitted. Finally, Corollary~\ref{cor:gap} on a block of size $m_i$ gives $(m_i/n)\gap(M_i)\ge(1/n)(1+1/m_i)$. Taking the minimum proves \eqref{eq:ordinalbound}.
\end{proof}

For disconnected posets, the original AST paper proves that $1-(n+2)/n^2$ occurs as an eigenvalue \cite{AST17}. Together with Corollary~\ref{cor:gap}, this places the gap in the interval $[(n+1)/n^2,(n+2)/n^2]$. The exact value throughout this class remains part of the spectral question.

\paragraph{AI disclosure.}
OpenAI GPT and Anthropic Claude assisted with proof development, review, and manuscript preparation. The author set the research directions, selected and revised the approaches, and takes responsibility for the arguments, references, and conclusions.

\appendix
\section{The transport entropy deficit}\label{app:deficit}

The proof of Theorem~\ref{thm:continuous} also decomposes its deficit into three nonnegative terms. Assume $\rho(F)=1$ and $0<c\le F\le C$. Let $A$ be the transport Hessian constructed in Section~\ref{sec:transport}. On a regular path write
\[
 r(u)=F(\Gamma(u)),\quad k(u)=A_{z(u)z(u)}(\Gamma(u)),\quad
 m=\int_0^1r(u)\,du,\quad a=\int_0^1k(u)\,du.
\]
Then $m>0$ and $0<a\le1$ almost surely.

Combining the pathwise Gibbs identity with $\det A=F$ gives
\begin{align}\label{eq:deficitidentity}
 n\Delta(F)-\Ent_\rho(F)
 ={}&\int F\log\frac{\prod_{z\in P}A_{zz}}{\det A}\dd\rho\notag\\
 &+n\E_{y,q}\left[mD\left(\frac rm\middle\Vert\frac ka\right)\right]
 -n\E_{y,q}[m\log a].
\end{align}
To justify the identity, rearrange \eqref{eq:gibbs} on each path:
\[
 \Ent_{du}(r)-\int r\log k
 =mD\left(\frac rm\middle\Vert\frac ka\right)-m\log a.
\]
The left side is integrable by Section~\ref{sec:transport}. The two terms on the right are nonnegative, so their expectations are separately finite. Averaging, multiplying by $n$, and adding the Hadamard deficit proves \eqref{eq:deficitidentity}. Each term is nonnegative: the determinant ratio is at least one, relative entropy is nonnegative, and $a\le1$.

The three losses have distinct meanings. The first measures the contribution of off-diagonal Hessian entries. The second compares the normalized input density along a path with the normalized diagonal Hessian. The third records the unused part of the unit Hessian-mass budget. Their sum is determined by the input entropy and the explicit conditional channel $H$.

For a centered discrete function $g$, put $B(t)=g(X(t))$ and $F_\epsilon=1+\epsilon B$. Using \eqref{eq:entropy-expansion} directly on the left side of \eqref{eq:deficitidentity}, together with \eqref{eq:varianceenergy}, gives
\begin{equation}\label{eq:deficitexpansion}
 n\Delta(F_\epsilon)-\Ent_\rho(F_\epsilon)
 =\frac{\epsilon^2}{2}
 \left[\frac{n^2}{n+1}\cE_M(g,g)-\Var_\pi(g)\right]
 +O_g(\epsilon^3).
\end{equation}

The conjectured bound $\gap(M)\ge(n+2)/n^2$ is therefore equivalent to the following strengthened quadratic inequality for functions that are constant on each ordering chamber:
\begin{equation}\label{eq:remaining}
 n V(B)-\Var_\rho(B)\ge\frac{1}{n+1}\Var_\rho(B),
 \qquad B(t)=g(X(t)).
\end{equation}
Theorem~\ref{thm:continuous} gives the same inequality with zero on the right. Linearizing the product densities in \eqref{eq:productequality} at $h=1$ gives equality in \eqref{eq:continuousvariance}. Thus the stronger estimate must use the restriction to chamber functions. Establishing \eqref{eq:remaining} for arbitrary posets would give the remaining finite-size spectral correction. The chamber entropy bound $\Ent_\rho(F)\le n(n+1)\Delta(F)/(n+2)$ would imply \eqref{eq:remaining} by linearization.

\section{Counting from uniform samples}\label{app:counting}

Write $e(P)=|\cL(P)|$. Combining the sampling theorem with the classical self-reduction for counting gives the following bound \cite{JVV86}. The reduction uses uniform distributions on intermediate posets. Banks, Garrabrant, Huber, and Perizzolo use a weighted TPA embedding, with a counting bound quadratic in $\log e(P)$ \cite{BGHP18}.

\begin{corollary}\label{cor:counting}
In the order-query model of Corollary~\ref{cor:sampling}, for $0<\varepsilon,\delta\le1/2$, there is a randomized algorithm returning $\widehat N$ such that
\[
 \Pr\bigl((1-\varepsilon)e(P)\le\widehat N\le(1+\varepsilon)e(P)\bigr)\ge1-\delta.
\]
After the same initial preprocessing, its running time is
\begin{equation}\label{eq:counting-time}
 O\!\left(n^4\varepsilon^{-2}\log^2 n\log\frac n\varepsilon\log\frac1\delta\right).
\end{equation}
\end{corollary}

\begin{proof}[Proof of Corollary~\ref{cor:counting}]
Fix a deterministic comparison-sorting algorithm, such as mergesort, with at most $L=n\lceil\log_2 n\rceil$ comparisons on every input. Its comparison tree has one permutation at each reachable leaf. Starting from $P_0=P$, follow a path in this tree. If a comparison is already determined by the current poset, follow that answer. Otherwise the two possible answers partition its linear extensions into two nonempty sets. Choose one answer and add the corresponding relation. After the last comparison the resulting poset is a total order.

Suppose the undetermined comparisons along the chosen path give posets $P_0,P_1,\ldots,P_\ell$, where $\ell\le L$ and $e(P_\ell)=1$. Write
\begin{equation}\label{eq:count-ratios}
 p_i=\frac{e(P_i)}{e(P_{i-1})},\qquad
 \prod_{i=1}^\ell p_i=\frac1{e(P)}.
\end{equation}
The probability $p_i$ is the chance that a uniform extension of $P_{i-1}$ satisfies the selected comparison. We choose the path so that every selected answer has probability bounded away from zero.

First analyze an ideal procedure using independent exact uniform samples. At each undetermined comparison, draw $b=\lceil18\log(128L)\rceil$ pilot samples from the current poset and select the empirical majority answer. For either fixed comparison, Hoeffding's inequality gives
\[
 \Pr\bigl(|\widehat p-p|>1/6\mid\text{previous pilot outcomes}\bigr)
 \le2e^{-b/18}\le\frac1{64L}.
\]
On the complementary event the chosen answer has true probability at least $1/3$. A union bound over the adaptively chosen comparisons therefore shows that, with probability at least $1-1/64$, the completed path satisfies $p_i\ge1/3$ for every $i$.

Freeze this path. For each $i$, use $s=\lceil4096L/\varepsilon^2\rceil$ fresh uniform samples from $\cL(P_{i-1})$ to estimate $p_i$ by the empirical frequency $\widehat p_i$. Conditional on the pilot stage these estimates are independent. Set $Z=\prod_i\widehat p_i$. Conditional on a path with all $p_i\ge1/3$, independence gives
\begin{align}\label{eq:count-variance}
 \E Z&=\prod_i p_i=\frac1{e(P)},\notag\\
 \frac{\Var(Z)}{(\E Z)^2}
 &=\prod_i\left(1+\frac{1-p_i}{sp_i}\right)-1
 \le e^{2L/s}-1
 \le\frac{4L}{s}
 \le\frac{\varepsilon^2}{1024}.
\end{align}
Chebyshev's inequality implies $|Z/\E Z-1|\le\varepsilon/4$ except with probability $1/64$. On this event $Z>0$, and the estimate $1/Z$ lies in $[(1-\varepsilon)e(P),(1+\varepsilon)e(P)]$. Define the output to be zero if $Z=0$. The ideal procedure consequently succeeds with probability at least $1-1/32$. If no undetermined comparison occurs, the poset already has one extension and we return one.

We now implement the procedure with Corollary~\ref{cor:sampling}. One repetition uses at most $T=L(b+s)$ sampling calls. Run every call with total variation error at most $\eta=1/(128T)$, using fresh randomness. Conditional on any past transcript, its output can be coupled to an exact uniform sample from the same current poset with disagreement probability at most $\eta$. Induction over the calls bounds the probability of any disagreement by $T\eta=1/128$. This applies to both the adaptive pilot stage and the estimation stage. Hence the implemented repetition fails with probability at most $1/32+1/128=5/128<1/8$.

Repeat independently an odd number $r\ge8\log(2/\delta)$ of times and return the median of the estimates. A Hoeffding bound for the independent failure indicators gives failure probability at most $e^{-r/4}\le\delta$. It remains to bound the running time.

We have $T=O(L^2\varepsilon^{-2})$ and $\log(1/\eta)=O(\log(n/\varepsilon))$. Each sampling call therefore costs $O(n^2\log(n/\varepsilon))$ operations. All intermediate posets can use the same transitive-closure table, updated in $O(n^2)$ time when a relation $a<b$ is added: insert every pair in $(\{a\}\cup\operatorname{pred}(a))\times(\{b\}\cup\operatorname{succ}(b))$. These are exactly the new reachable pairs. An initial extension for each current poset can be found by a topological sort in $O(n^2)$ time. The path can be stored as a list of added relations and reconstructed for the estimation stage. Thus maintaining the posets costs $O(Ln^2)$ per repetition, which is smaller than the sampling cost.

The frequencies have denominator $s$, so each reciprocal product can be stored exactly using $O(L\log s)$ bits. Ordinary integer arithmetic forms these products in $O(L^2\log^2 s)$ bit operations; a linear-time median selection uses $O(r)$ comparisons of numbers of this size. These costs are absorbed by the sampling bound for $0<\varepsilon\le1/2$. Multiplying the sampling cost by $r$ and using $L=O(n\log n)$ proves \eqref{eq:counting-time}.
\end{proof}


\begin{thebibliography}{CCGP25}
\small
\setlength{\itemsep}{3pt}

\bibitem[AKS14a]{AKS14a}
Arvind Ayyer, Steven Klee, and Anne Schilling.
\newblock Combinatorial Markov chains on linear extensions.
\newblock \emph{Journal of Algebraic Combinatorics} 39 (2014), 853--881.
\newblock \href{https://doi.org/10.1007/s10801-013-0470-9}{doi:10.1007/s10801-013-0470-9}.

\bibitem[AKS14b]{AKS14b}
Arvind Ayyer, Steven Klee, and Anne Schilling.
\newblock Markov chains for promotion operators.
\newblock In \emph{Algebraic Monoids, Group Embeddings, and Algebraic Combinatorics}, Fields Institute Communications 71, Springer, 2014, 285--304.
\newblock \href{https://doi.org/10.1007/978-1-4939-0938-4_13}{doi:10.1007/978-1-4939-0938-4\_13}.

\bibitem[AST17]{AST17}
Arvind Ayyer, Anne Schilling, and Nicolas M. Thi\'ery.
\newblock Spectral gap for random-to-random shuffling on linear extensions.
\newblock \emph{Experimental Mathematics} 26 (2017), no.~1, 22--30.
\newblock \href{https://doi.org/10.1080/10586458.2015.1107868}{doi:10.1080/10586458.2015.1107868}.

\bibitem[Bar98]{Bar98}
Franck Barthe.
\newblock On a reverse form of the Brascamp--Lieb inequality.
\newblock \emph{Inventiones Mathematicae} 134 (1998), no.~2, 335--361.
\newblock \href{https://doi.org/10.1007/s002220050267}{doi:10.1007/s002220050267}.

\bibitem[BD99]{BD99}
Russ Bubley and Martin Dyer.
\newblock Faster random generation of linear extensions.
\newblock \emph{Discrete Mathematics} 201 (1999), no.~1--3, 81--88.
\newblock \href{https://doi.org/10.1016/S0012-365X(98)00333-1}{doi:10.1016/S0012-365X(98)00333-1}.

\bibitem[BGHP18]{BGHP18}
Jacqueline Banks, Scott Garrabrant, Mark L. Huber, and Anne Perizzolo.
\newblock Using TPA to count linear extensions.
\newblock \emph{Journal of Discrete Algorithms} 51 (2018), 1--11.
\newblock \href{https://doi.org/10.1016/j.jda.2018.04.001}{doi:10.1016/j.jda.2018.04.001}.

\bibitem[BN19]{BN19}
Megan Bernstein and Evita Nestoridi.
\newblock Cutoff for random to random card shuffle.
\newblock \emph{The Annals of Probability} 47 (2019), no.~5, 3303--3320.
\newblock \href{https://doi.org/10.1214/19-AOP1340}{doi:10.1214/19-AOP1340}.

\bibitem[Bre91]{Bre91}
Yann Brenier.
\newblock Polar factorization and monotone rearrangement of vector-valued functions.
\newblock \emph{Communications on Pure and Applied Mathematics} 44 (1991), no.~4, 375--417.
\newblock \href{https://doi.org/10.1002/cpa.3160440402}{doi:10.1002/cpa.3160440402}.

\bibitem[CCGP25]{CCGP25}
Pietro Caputo, Zongchen Chen, Yuzhou Gu, and Yury Polyanskiy.
\newblock Entropy contractions in Markov chains: Half-step, full-step and continuous-time.
\newblock \emph{Electronic Journal of Probability} 30 (2025).
\newblock \href{https://doi.org/10.1214/25-EJP1372}{doi:10.1214/25-EJP1372}.

\bibitem[CL10]{CL10}
Dario Cordero-Erausquin and Michel Ledoux.
\newblock The geometry of Euclidean convolution inequalities and entropy.
\newblock \emph{Proceedings of the American Mathematical Society} 138 (2010), no.~8, 2755--2769.
\newblock \href{https://doi.org/10.1090/S0002-9939-10-10304-9}{doi:10.1090/S0002-9939-10-10304-9}.

\bibitem[CMT15]{CMT15}
Pietro Caputo, Georg Menz, and Prasad Tetali.
\newblock Approximate tensorization of entropy at high temperature.
\newblock \emph{Annales de la Facult\'e des Sciences de Toulouse. Math\'ematiques}, S\'erie~6, 24 (2015), no.~4, 691--716.
\newblock \href{https://doi.org/10.5802/afst.1460}{doi:10.5802/afst.1460}.

\bibitem[CP25]{CP25}
Swee Hong Chan and Igor Pak.
\newblock Linear extensions of finite posets.
\newblock \emph{EMS Surveys in Mathematical Sciences} (2025), published online first.
\newblock \href{https://doi.org/10.4171/EMSS/97}{doi:10.4171/EMSS/97}.

\bibitem[DPF14]{DPF14}
Guido De Philippis and Alessio Figalli.
\newblock The Monge--Amp\`ere equation and its link to optimal transportation.
\newblock \emph{Bulletin of the American Mathematical Society} 51 (2014), no.~4, 527--580.
\newblock \href{https://doi.org/10.1090/S0273-0979-2014-01459-4}{doi:10.1090/S0273-0979-2014-01459-4}.

\bibitem[DS18]{DS18}
A. B. Dieker and F. V. Saliola.
\newblock Spectral analysis of random-to-random Markov chains.
\newblock \emph{Advances in Mathematics} 323 (2018), 427--485.
\newblock \href{https://doi.org/10.1016/j.aim.2017.10.034}{doi:10.1016/j.aim.2017.10.034}.

\bibitem[Han78]{Han78}
Te Sun Han.
\newblock Nonnegative entropy measures of multivariate symmetric correlations.
\newblock \emph{Information and Control} 36 (1978), no.~2, 133--156.
\newblock \href{https://doi.org/10.1016/S0019-9958(78)90275-9}{doi:10.1016/S0019-9958(78)90275-9}.

\bibitem[Hub06]{Hub06}
Mark Huber.
\newblock Fast perfect sampling from linear extensions.
\newblock \emph{Discrete Mathematics} 306 (2006), no.~4, 420--428.
\newblock \href{https://doi.org/10.1016/j.disc.2006.01.003}{doi:10.1016/j.disc.2006.01.003}.

\bibitem[Hub25]{Hub25}
Mark Huber.
\newblock Generating uniform linear extensions using few random bits.
\newblock Preprint (2025), \href{https://arxiv.org/abs/2506.14725}{arXiv:2506.14725}.

\bibitem[JVV86]{JVV86}
Mark R. Jerrum, Leslie G. Valiant, and Vijay V. Vazirani.
\newblock Random generation of combinatorial structures from a uniform distribution.
\newblock \emph{Theoretical Computer Science} 43 (1986), 169--188.
\newblock \href{https://doi.org/10.1016/0304-3975(86)90174-X}{doi:10.1016/0304-3975(86)90174-X}.

\bibitem[LLO25]{LLO25}
Jonathan Leake, Kasper Lindberg, and Shayan Oveis Gharan.
\newblock Optimal trickle-down theorems for path complexes via $\mathcal C$-Lorentzian polynomials with applications to sampling and log-concave sequences.
\newblock In \emph{Proceedings of the 66th IEEE Symposium on Foundations of Computer Science (FOCS)}, 2025, 1857--1870.
\newblock \href{https://doi.org/10.1109/FOCS63196.2025.00097}{doi:10.1109/FOCS63196.2025.00097}.

\bibitem[McC97]{McC97}
Robert J. McCann.
\newblock A convexity principle for interacting gases.
\newblock \emph{Advances in Mathematics} 128 (1997), no.~1, 153--179.
\newblock \href{https://doi.org/10.1006/aima.1997.1634}{doi:10.1006/aima.1997.1634}.

\bibitem[PS18]{PS18}
Svetlana Poznanovi\'c and Kara Stasikelis.
\newblock Properties of the promotion Markov chain on linear extensions.
\newblock \emph{Journal of Algebraic Combinatorics} 47 (2018), 505--528.
\newblock \href{https://doi.org/10.1007/s10801-017-0783-1}{doi:10.1007/s10801-017-0783-1}.

\bibitem[RS22]{RS22}
John Rhodes and Anne Schilling.
\newblock Upper bounds on mixing time of finite Markov chains.
\newblock \emph{SIAM Journal on Discrete Mathematics} 36 (2022), no.~4, 3031--3057.
\newblock \href{https://doi.org/10.1137/21M1421994}{doi:10.1137/21M1421994}.

\bibitem[Sta86]{Sta86}
Richard P. Stanley.
\newblock Two poset polytopes.
\newblock \emph{Discrete \& Computational Geometry} 1 (1986), 9--23.
\newblock \href{https://doi.org/10.1007/BF02187680}{doi:10.1007/BF02187680}.

\bibitem[Wil04]{Wil04}
David B. Wilson.
\newblock Mixing times of lozenge tiling and card shuffling Markov chains.
\newblock \emph{The Annals of Applied Probability} 14 (2004), no.~1, 274--325.
\newblock \href{https://doi.org/10.1214/aoap/1075828054}{doi:10.1214/aoap/1075828054}.

\end{thebibliography}
\end{document}